\documentclass[12pt,aps,draft,showpacs]{revtex4}
\usepackage{bm}
\usepackage{epsfig}
\begin{document}
\title{A position and a time for the photon}
\author{J. Le\'on}\email{leon@imaff.cfmac.csic.es}
\affiliation{Instituto de Matem\'aticas y F\'{\i}sica
Fundamental, CSIC\\
Serrano 113-bis, 28006 MADRID, Spain}

\date{\today}

\begin{abstract}
This paper gives a constructive answer to the question whether  photon states can contain or not, and to what extent, the readings of rulers and clocks. The paper first shows explicitly that, along with the momentum  representation, there is room in the one photon Hilbert space for an alternative position representation. This is made possible by the existence of a self-adjoint, involutive, position operator conjugate to the momentum operator~\cite{hawtona}. Position and momenta are shown to satisfy the Heisenberg-Weyl quantization rules in the helicity basis, which is analyzed anew from this point of view. The paper then turns to the photon's time of arrival. By picking an appropriate photon Hamiltonian - using Maxwell equations as the photon Schr\"{o}dinger equation - a conjugate time of arrival operator is built. Its interpretation, including the probability densities for the instant of arrival (at arbitrary points of 3-D space) of photon states with different helicities coming from arbitrary places, is discussed. 

\end{abstract}

\pacs{03.65.-w, 14.70.Bh, 42.50.Dv} \maketitle
\section*{Introduction}
Since the advent of Special Relativity, light propagation 
has played a
very special role among the variety of natural phenomena
studied by Physics. In conjunction with rods and clocks,
light signals were a main component in the construction of
the -then- new theory. The universality of  $c$, the speed
of light, set the new standard commensurate to the breadth
of relativistic physics. While light was understood at the
time as a manifestation of electromagnetic waves, the
interaction of matter and radiation demanded a quantum leap:
the photon. This is the place to recall the mismatch between the prominent role played in relativity by the arrival times of light signals and the position of their group centers and wave fronts, and the lack of the corresponding operators in quantum physics~\cite{pauli,pryce,newton}. The reason of this deficiency of the quantum theory~\cite{wightman} can be traced back to the fact that the photon does not have a complete set of helicity states; the state with helicity $\lambda=0$ is missing. This paper aims to give a constructive answer to the long debated question whether the photon state can contain or no, and to what
extent, the readings of rulers and clocks, i.e.  the where's
and when's of the photon. In view of the recent experimental
advances, an appropriate formalism is necessary for a
meaningful discussion of the propagation and arrival of
light signals, mainly if these arrive photon by photon. The literature on this question is abundant and only a part of it is 
in the bibliography at the end of this paper. The reader is referred to~\cite{scully} for a concise and clear introduction to this topic.

Electromagnetic waves propagate at the speed of light in
vacuum. This basic tenet of relativity, has been challenged
recently both from the experimental and the theoretical
front. It is remarkable that the theoretical part of this
question was settled long
time ago by Sommerfeld~\cite{sommerfeld} and Brillouin~\cite{brillouin}, who
demonstrated that the electromagnetic wave fronts always
move with speed $c$. To my knowledge, the cases where the
group velocity in vacuum was found larger than $c$ where always
produced by an incorrect identification of the signal
center. Superluminal propagation in vacuum turned always
to be a geometric
artifact of the wave form and not the group speed of the
wave packet. The main point to highlight here is
the absence of foundational problems associated to the
definitions of the group position and time of arrival of classical
e.m. waves. It may be difficult to ascertain these questions
in a given situation, but only from the technical point of
view, not as a matter of principle.

Electromagnetic waves are composed of elementary quanta:
photons.
Like the rest of elementary particles, these are
mathematically identified as elementary systems under the
Poincare group~\cite{wigner,bargmann}. Namely, all the possible sets of values
that the
dynamical properties of the system can attain are connected
 by
transformations
of the group to an arbitrary, fixed, set of values, taken
as standard.
The different classes of elementary systems are identified
by the
values taken by the Casimir operators of the group. Photons
 are
characterized by lightlike momenta $p^2=(p^0)^2-\bm{p}^2=0$
 and a helicity value that is
either $\lambda=+1$ or $\lambda=-1$, there are thus two
irreducible
representations
(1,0) and (0,1) (in the standard notation $(j,j')$), that are
 combined
into an irreducible entity by
parity (1,0)$\stackrel{\cal P}{\rightarrow}$(0,1). Finally,
 photons
transform under the representation $(1,0)+(0,1)$ of the
complete
Poincare group.

The properties of photons are thus given by the eigenvalues
of the subsets of commuting generators of the Poincare group
in that representation. There are two widely used
alternative subsets: a). $\{\bm{p}, \,\lambda\}$, that
describe plane waves, suited for the description of initial
and final scattering states, free flying photons etc, and
b). $\{p^0,j,j_z,parity\}$,
 that describe spherical photon waves
that simplify the analysis of radiation, and in general,
of those systems where there is a singled out space point.

The rapid pace and depth of the current advances in the
manipulation of photon states are demanding a parallel
progress on the theoretical side. In some cases it is
necessary to describe accurately the behavior of photons
that are in the near region. In others, their space time
distributions in regions close to the microscopic realm, for
states containing just one photon or a few of them, in states
possibly entangled that spread over these regions, etc.
Current experimental set ups include a variety of very,
sensitive single photon devices. Even more, many of the most
exciting results have been obtained experimenting with these
highly nonclassical light states. On its side, the
theoretical framework should be able to address simple
questions like the time of arrival of a photon at the place
where some device is located, the relation between the
orbital angular momentum of the photon and its position at
some time, etc.  It is necessary at least to give a meaning
to these questions, something whose very possibility has
been the subject of debate~\cite{pauli,landaua,landaub,galindo,amrein,rosewarne} since the early times of
Quantum Mechanics. I shall avoid here repeating the well
known pros and cons for the wave function of the photon~\cite{cooka,cookb,inagaki,iwoa,iwob,sipe} and
whether it represents the probability amplitude for finding
the photon in a given region of space (technically a Borel set $\Delta \in \bm{R}^3$). Instead, I shall trace
the problem back by an alternative route to non-relativistic
quantum mechanics (NRQM) where the vanishing photon mass
($M=0$) prevents the definition of a position operator
$\hat{\bm{q}}$ for the photon.

Let me first give some support to the above assertion. The
dynamical properties of an elementary object, i.e. a free
particle, in non-relativistic quantum theory can be traced
back to its behavior under the Galileo group~\cite{jordan}.
In particular the position operator $\hat{\bm{q}}$ is
related to the Galileo boost operator $\hat{\bm{G}}$ by
$\hat{G}_i=M \hat{q}_i$, where $M$ is the mass of the
particle. For sure, it can be defined directly by $\hat{q}_i
|\bm{x}\rangle = x_i|\bm{x}\rangle$, so that its components
are commuting, its  eigenvalues real numbers, and its
eigenfunctions Dirac delta functions. On the other hand,
this definition has to be compatible with the fact that the
momentum operator $\hat{\bm{p}}$ generates translations in
configuration space, $[\hat{q}_i,\hat{p}_j]=i \hbar$.
Accordingly in the momentum representation, more appropriate
for a discussion of the Galileo group generators, the
position operator acts as a derivative on the momenta
$\hat{q}_i= i\hbar
\partial/\partial p_i$; if $\hat{p}_i$ is the basic,
multiplicative operator of the representation, then
$\hat{q}_i$ is derived from it. In other words, particle
position emerges from the Galilean properties of
the particle, specifically from its behavior under space
translations.

As a consequence of the above, one may wonder that if in
NRQM a massless system can not have a properly defined
position operator, then: Why expect the opposite in RQM?
 The contenders in the debate about the
photon wave function could be aware or not about the lack of
a non-relativistic $\hat{\bm{q}}$ for the photon. In any
case, they were undoubtedly influenced by their
consequences, and induced to give up in the search of a
relativistic $\hat{\bm{q}}$ for $M=0$. The fact that the
photon little group is $E(2)$ instead of $SU(2)$ added the
filling drop to the confusion. In some studies of the
position operator for relativistic systems,  photon
transversality was identified as the obstacle to the existence of a photon
position operator~\cite{newton,wightman}. Other analysis~\cite{pryce,iwob} pointed to the lack of commutativity of its components (or better, of the Lorentz boost components) as the reason for the lack of an
appropriate $\hat{\bm{q}}$. Finally, M. Hawton found out~\cite{hawtona,hawtonb,hawtonc} very recently a
self-adjoint position operator with the required commutation
rules, that operates in the photon Hilbert space as a
physically acceptable position should do. This opens up the possibility of analyzing the question of photon localizability from a new and more powerful setting: A good position operator should come endowed with a probabilistic interpretation, which is the  tool suited to the analysis of localizability~\cite{wightman,rosewarne,galindo,amrein}. On the other hand, the most recent theoretical analysis~\cite{adlard,iwoc} indicate that individual photons can be localized in regions much smaller than previously thought, with entanglement (instead of wavelength) as the physical ruler~\cite{chan} in the experiment~\cite{kurtsiefer}. This increases the value of the Hawton construct and the urge to explore its consequences.

In this paper I shall analyze the properties of the photon
states in Minkowski space-time.  On the way, we shall devise
several tools necessary to this end. The photon Hilbert
space will be introduced in Sect. II, where we also present
two alternative conjugate representations and show the
probabilistic interpretation that they can be given. Sect.
III is devoted to the explicit construction of the Hawton
position operator in these representations. We revise the
properties of the operator in the helicity basis and
conclude that it qualifies as a good position operator.
Finally, we complete the standard picture of one photon
states with additional position dependent information.
Conversely, Sect. IV is devoted to the time dependent
information. We first consider Maxwell equations as the
photon Sch\"odinger equation and solve the time evolution
associated to it. Then, we set up the formalism for
analyzing the photon time of arrival at an arbitrary point
of space. It will be necessary to split the Hilbert space
into two subspaces putting the eventually detected states in
one of them. In it we build  the  sought time operator,
obtain its eigenfunctions and give the positive operator
valued measure that permits a probabilistic analysis of the
times of arrival of a photon in a given state at an
arbitrary space position. The paper ends in Sect. V with a
summary of the results, some considerations on their
meaning, and indications for the use of the formalism in the
analysis of some current experiments.

\section*{One photon Hilbert space}

We shall work in the Coulomb gauge i.e. within the Hilbert 
space $\cal{H}$ of one
particle transverse states
$\tilde{A}_i(\bm{p}),\; i=1,2,3$ defined on the forward
light cone $(p^0=|\bm{p}|)$ (note  the symbol $\tilde{}$ , it
is a label for transversality).  The scalar product in
$\cal{H}$ is defined  as
\begin{equation}
(A,A') = \int d\sigma(\bm{p}) \tilde{A}^*_i (\bm{p})
\tilde{A'}_i (\bm{p}) \label{p2}
\end{equation}
where $d\sigma(\bm{p}) =d^3p/2 |\bm{p}|$ is the measure on
the light cone. An arbitrary photon state $\bm{A}\in\cal{H}$
can be written as
\begin{equation}
\tilde{A}_i (\bm{x}) = (2 \pi)^{-3/2} \int \frac{d^3
p}{\sqrt{2 |\bm{p}|}} \; e^{i \bm{p}\bm{x}}\; \tilde{A}_i
(\bm{p}) \label{p1}
\end{equation}
 Note that
$\tilde{A}_i (\bm{x})$ is solenoidal so that, in spite of
being $(A,A') = \int d^3 x \tilde{A}^*_i (\bm{x})
\tilde{A'}_i (\bm{x})$, $\bm{x}$ does not qualify for a good
position (in poor words: $x_i \tilde{A}(\bm{x})$ is not solenoidal, so $x_i$ takes $\tilde{A}$ out of the Hilbert space).

Due to Poincare invariance (that we are taking fully into account, in spite of the non covariant notation), the
helicity $W$ is a good quantum number. In fact,
$\hat{\bm{p}}$ and $\hat{W}=\bm{S}
\widehat{(\bm{p}/|\bm{p}|)}$ -- where
$S^a_{ij}=-i\epsilon_{aij}$ are the spin-1 matrices -- form
a complete set of commuting operators on the mass shell. In
the following, I shall work in momentum space where the
momentum operator $\hat{\bm{p}}$ is simply represented by
the vector variable $\bm{p}=(p_1,p_2,p_3)$. An arbitrary
operator $\hat{G}$ function of the momentum, that commutes
with the helicity $[ \hat{G},\hat{W}]=0$, shall have
eigenfunctions $\tilde{V}^i_{G,\lambda}(\bm{p})$ satisfying
\begin{eqnarray}
(\hat{p}^a \,\tilde{V}_{G,\lambda})^i (\bm{p}) = p^a \,
\tilde{V}^i_{G,\lambda}(\bm{p}), \; & (\hat{W}
\,\tilde{V}_{G,\lambda})^i (\bm{p}) = \lambda \,
\tilde{V}^i_{G,\lambda}(\bm{p}),\; & (\hat{G} \;\;
\tilde{V}_{G,\lambda})^i (\bm{p})= G\,
\tilde{V}^i_{G,\lambda}(\bm{p})\; \label{p3}
\end{eqnarray}
The second of these equations implies that
$\tilde{V}^i_{G,\lambda}(\bm{p})\propto \epsilon^i (\bm{p},
\lambda)$, where $\bm{\epsilon} (\bm{p}, +1)$
($\bm{\epsilon} (\bm{p}, -1)$) are the right (left)
polarization vectors for momentum $\bm{ p}$. There is some
arbitrariness in their definition,  the convention of
Ref.~\cite{weinberga} adopted here is $\epsilon_i (\bm{p},
\lambda)=D_{ij}[\Lambda_{\bm{p}\leftarrow \bm{k}}]\,
 \epsilon_j (\lambda)$, for $\bm{p}=\Lambda \, \bm{k}$
and $\bm{k}=(0,0,|\Lambda^{-1}\bm{p}|)$, with
$\bm{\epsilon}(\lambda)$ such that $S_3
\,\bm{\epsilon}(\lambda)=\lambda \bm{\epsilon}(\lambda)$. In
spherical coordinates where $\bm{p}= (k,\theta,\varphi)$ and
$\bm{k}=(0,0,k)$ -- i.e. taking $\Lambda$ as a pure rotation
for simplicity -- the polarization vectors are given as
linear combinations of the unitary vectors
$\bm{e}(\bm{p},\sigma),\; \sigma=\theta,\varphi,k$ (or
$\sigma=1,2,3$):
$$\bm{e}(\bm{p},k)=(\partial \bm{ p}/ \partial k),\,
\bm{e}(\bm{p},\theta)=(1/k)\, (\partial \bm{p}/
\partial \theta),\, \bm{e}(\bm{p},\varphi)=
(1/k\sin\theta)\, (\partial
\bm{p}/\partial \varphi)$$ in the standard form:
\begin{equation}
 \epsilon^i (\bm{p}, \lambda)= -
\frac{\lambda}{\sqrt{2}}\{e^i(\bm{p},\theta)+i \lambda \,
e^i(\bm{p},\varphi)\}\;\;\,\, \lambda=\pm 1 \label{p4}
\end{equation}

The representations of the Poincare group for the photon
states are the (1,0) and (0,1), that correspond to
$\lambda=-1$ (left polarization) and  $\lambda=1$ (right
polarization) respectively. They are combined in the direct
sum $(1,0) \oplus (0,1)$ that becomes irreducible when
parity is included in the group. No other states
exist (see for instance~\cite{weinberga} page 69) because the little group is not
semi-simple. In particular, the polarizations can not be
taken as four-vectors. Even if we form the object
$\epsilon^\mu  (p, \lambda)=(\epsilon^0 (p,
\lambda),\bm{\epsilon} (p, \lambda))$, it not only has to
satisfy in every reference frame $ p_\mu \epsilon^\mu  (p,
\lambda)=0$, but also $\epsilon^0  (p, \lambda)=0$, or any
other condition chosen to fix the gauge. In short: in
spite of its four-dimensional appearance, the polarizations
do not transform as
 four-vectors under Lorentz transformations, but as a sort of
connections~\cite{weinbergb,weinbergc}:
\begin{equation}
\Lambda_\nu^\mu \epsilon^\nu (\Lambda\, p,\lambda)=\exp\{i
\lambda \Theta(p,\Lambda)\} \epsilon^\mu (p,\lambda) + p^\mu
\Omega(p,\lambda;\Lambda)\label{p4a}
\end{equation}
where $\Theta(p,\Lambda)$ is the Wigner rotation angle, and
$\Omega(p,\lambda;\Lambda)$ is a gauge transformation fixed
by the gauge condition ($\epsilon^0=0)$. Of course, the
photons can be given a
 covariant (tensor) representation by means of the gauge invariant
antisymmetric tensor $F_{\alpha \beta}(p,\lambda) \propto i
\epsilon_{\alpha \beta \mu \nu} (p^\mu
\epsilon^\nu(p,\lambda)-p^\nu \epsilon^\mu(p,\lambda))$,
that gives rise to the electric and magnetic fields. With
these caveats in mind, I shall continue to use the three
dimensional notation to work in the Coulomb gauge in the
following, turning to~\cite{weinbergb,weinbergc} for questions of
Poincare covariance or gauge invariance.

By construction,
\begin{equation}
(\bm{S}\cdot \bm{e}(\bm{p},k))_{ij} =(S_k)_{ij},\;\;\;
\mbox{and}\;\;\; (S_k)_{ij} \epsilon_j (\bm{p}, \lambda)=
\lambda \epsilon_i (\bm{p}, \lambda),\;\; \lambda=\pm 1
\label{p5}
\end{equation}
We can now write  $\tilde{V}^i_{G,\lambda}(\bm{p}) =g_G
(\bm{p})\, \epsilon^i (\bm{p}, \lambda)$, where $g_G
(\bm{p})$ is a $G$ dependent function of $\bm{p}$ to be
determined in each case. By these definitions, these
functions span the transverse subspace orthogonal to
$\bm{p}$, and are eigenfunctions of the helicity with
eigenvalue $\lambda$.

\subsection*{Momentum and position representations}
In the trivial case of the momentum operator
$\hat{\bm{G}}=\bm{p}$ where
$\tilde{V}^i_{\bm{p},\lambda}(\bm{p'}) =g_{\bm{p}}
(\bm{p'})\, \epsilon^i (\bm{p'}, \lambda) $, it is
straightforward to obtain $g_{\bm{p}}$. We want
$(V_{\bm{p}_1,\lambda_1},V_{\bm{p}_2,\lambda_2})=\delta_{\lambda_1
\lambda_2} \delta^{(3)}(\bm{ p}_1-\bm{  p}_2)$, but we only have
\begin{equation}
(V_{\bm{p}_1,\lambda_1},V_{\bm{p}_2,\lambda_2}) =\int
d\sigma(\bm{p}){\tilde{V}}^{*i}_{\bm{p}_1,\lambda_1}
(\bm{p}) \tilde{V}^i_{\bm{p}_2,\lambda_2}
(\bm{p})=\delta_{\lambda_1 \lambda_2} \int
d\sigma(\bm{p})g^*_{\bm{p}_1} (\bm{p})g_{\bm{p}_2}
(\bm{p})\label{p6}
\end{equation}
where we used that $\bm{\epsilon}^* (\bm{p}, \lambda)
\bm{\epsilon} (\bm{p}, \lambda')=\delta_{\lambda \lambda'}$.
This requires the value $g_{\bm{p}}(\bm{p}')=\sqrt{2
|\bm{p}'|} \delta^{(3)}(\bm{ p}'-\bm{ p})$ for $g_{\bm{p}}$,
so that, apart from phases, the eigenfunctions have to be
\begin{eqnarray}
\tilde{V}^i_{\bm{p},\lambda}(\bm{p}') =\sqrt{2 |\bm{p}'|}\,
\epsilon^i (\bm{p}', \lambda)\, \delta^{(3)}(\bm{  p}'-\bm{
p}),\;\;\; & \mbox{and}\;\;\; &
\tilde{V}^i_{\bm{p},\lambda}(x) =\frac{1}{(2\pi)^{3/2}} \,
\epsilon^i (\bm{p}, \lambda)\, e^{i \bm{p}\bm{x}}\label{p8}
\end{eqnarray}
Note that the second equation in (\ref{p8}) comes from the first one by straight
application of Eq (\ref{p1}).

The next step is to assume the existence of a position, namely a
vector operator
$\hat{\bm{G}}=\hat{\bm{q}}=(\hat{q_1},\hat{q_2},\hat{q_3})$,
with simultaneous
 eigenfunctions in $\cal{H}$ that can be given as
 $\tilde{V}^i_{\bm{q},\lambda}(\bm{p})
=g_{\bm{q}} (\bm{p})\, \epsilon^i (\bm{p}, \lambda) $.  For
this to be possible it is necessary that:
\begin{itemize}
\item[i.] Position and helicity commute, that is
$[\hat{q}^i,\hat{W}]=0$.
\item[ii.] The three components are in involution, that is
$[\hat{q}^i,\hat{q}^j]=0$.
\end{itemize}
To these conditions we have to add that positions and
momenta be canonically conjugate operators:
$[\hat{q}^i,\hat{p}^j]=i \delta^{ij}$. Finally, it will be
necessary to check whether $\hat{\bm{q}}$ gives rise to a
probabilistic interpretation or not.

 We shall return to the explicit construction of the position
 operator later on but, for the time being, we assume its
 eigenfunctions exist in the transverse Hilbert space of the photon
 and form an orthogonal
 and complete set. We first explore the implications of the
orthogonality. A direct computation using
$\tilde{V}^i_{\bm{q},\lambda}(\bm{p}) =g_{\bm{q}} (\bm{p})\,
 \epsilon^i(\bm{p}, \lambda) $gives
\begin{equation}
(V_{\bm{q}_1,\lambda_1},V_{\bm{q}_2,\lambda_2})=\delta_{\lambda_1
\lambda_2} \int d\sigma(\bm{p}) g^*_{\bm{q}_1} (\bm{p})
g_{\bm{q}_2} (\bm{p}) \label{p9}
\end{equation}
but, as in the case of the momentum operator, we would like
to have  $(V_{\bm{q}_1,\lambda_1},V_{\bm{q}_2,\lambda_2}) =
\delta_{\lambda_1 \lambda_2} \delta^{(3)}(\bm{ q}_1-\bm{
q}_2)$.
 This requires  $g_{\bm{q}}(\bm{p})= (2 \pi)^{-3/2} \,\sqrt{2
|\bm{p}|}\,\, e^{-i \bm{p}\bm{q}}$. We can now give the
position eigenfunctions in both, momentum and coordinate
representations:
\begin{eqnarray}
\tilde{V}^i_{\bm{q},\lambda}(\bm{p}) = (2 \pi)^{-3/2}\,
\sqrt{2 |\bm{p}|}\, \epsilon^i (\bm{p}, \lambda)\, e^{-i
\bm{p}\bm{q}} \; &\mbox{and} \;
&\tilde{V}^i_{\bm{q},\lambda}(\bm{x})=(2\pi)^{-3} \int d^3 p\,
\epsilon^i (\bm{p}, \lambda)\, e^{i
\bm{p}(\bm{x}-\bm{q})}\label{p11}
\end{eqnarray}
It is clear from the last equation that $\bm{x}$ does not
correspond to the position eigenvalue $\bm{q}$. This
mismatch is due to the presence in the integrand of the {\em
rhs} of  momentum dependent polarization vectors that link
momentum with spin.This equation summarizes much of the
troublesome nature of the coordinate space wave function of
the photon.

The above eigenfunctions form  complete sets. By
straightforward calculation using $\sum_\lambda
\epsilon^{*i} (\bm{p},\lambda) \epsilon^j
(\bm{p},\lambda)=\delta^{ij}-p^i p^j/|\bm{p}|^2 =
\delta^{ij}_\bot (\bm{p})$, we get the decompositions of the
identity
\begin{eqnarray}
\prod (\bm{p}_1,\bm{p}_2) =&\sum_\lambda \int
d\sigma(\bm{p}) \tilde{V}^i_{\bm{p},\lambda} (\bm{p}_1)
\tilde{V}^{j*}_{\bm{p},\lambda} (\bm{p}_2)&=
\delta^{ij}_\bot (\bm{p}_1)
\delta^{(3)}(\bm{ p}_1-\bm{ p}_2) \label{p11a}\\
\bigwedge (\bm{p}_1,\bm{p}_2) =& \sum_\lambda \int d^3 q
\tilde{V}^i_{\bm{q},\lambda} (\bm{p}_1)
\tilde{V}^{j*}_{\bm{q},\lambda} (\bm{p}_2)&= 2 |\bm{p}_1 |
\delta^{ij}_\bot (\bm{p}_1) \delta^{(3)}(\bm{ p}_1-\bm{
p}_2) \label{p11b}
\end{eqnarray}

We have arrived at two alternative complete sets of
commuting operators ({\sl momentum, helicity}) and ({\sl
position, helicity}), whose simultaneous eigenfunctions form
alternative bases of the Hilbert space $\cal{H}$ of the
massless $(1,0) \oplus (0,1)$ representation of the Poincare
group. We recall again the difference between position and
coordinate, and refer the reader back to Eq.(\ref{p11}) for
a check. Given an arbitrary state $\Psi$ in $\cal{H}$, we
could employ Dirac bracket notation to denote its components
in both representations
\begin{eqnarray}
\langle \bm{p} \lambda|\Psi\rangle=&\int
d\sigma(\bm{p}')\tilde{V}^{i*}_{\bm{p},\lambda} (\bm{p}')
\tilde{\Psi}^i (\bm{p}')=&
\frac{1}{\sqrt{2\,|\bm{p}|}}\epsilon^{i*} (\bm{p},\lambda)
\tilde{\Psi}^i
(\bm{p}) \label{p11m}\\
\langle \bm{q} \lambda|\Psi\rangle=&\int
d\sigma(\bm{p}')\tilde{V}^{i*}_{\bm{q},\lambda} (\bm{p}')
\tilde{\Psi}^i (\bm{p}')=& (2\pi)^{-3/2} \int d^3 p \,\,e^{i
\bm{p}\bm{q}}\,\, \langle \bm{p} \lambda|\Psi \rangle
\label{p11n}
\end{eqnarray}
The relation (\ref{p11n}) is the standard one between
position and momentum representations in quantum mechanics.
In fact,
\begin{equation}
\langle \bm{q} \lambda | \bm{p} \lambda' \rangle= (2
\pi)^{-3/2}\,\,\delta_{\lambda \lambda'} \exp(i \bm{p}
\bm{q}),\; \langle \bm{p} \lambda | \bm{p}' \lambda'
\rangle=\delta_{\lambda \lambda'} \delta^{(3)}(\bm{ p}-\bm{
p'}),\; \langle \bm{q} \lambda | \bm{q}' \lambda'
\rangle=\delta_{\lambda \lambda'} \delta^{(3)}(\bm{ q}-\bm{
q'}) \label{p11e}
\end{equation}
Note also the scalar product in both representations:
\begin{equation}
\langle \psi | \psi \rangle=\sum_\lambda \int d^3 p\,
|\langle \bm{p} \lambda|\Psi \rangle|^2 = \sum_\lambda \int
d^3 q\, |\langle \bm{q} \lambda|\Psi \rangle|^2 \label{p11z}
\end{equation}
The absence of polarization vectors in the integrand in the
{\sl rhs} of (\ref{p11n}) eliminates the
obstructions~\cite{wightman,galindo,amrein} that prevent from
considering
\begin{equation}
P_\psi (\Delta,\lambda)= \int_\Delta d^3 q |\langle \bm{q} 
\lambda|\Psi \rangle|^2 \label{p11y}
\end{equation}
as the probability for finding a photon of helicity
$\lambda$ in the Borel set $\Delta \in \mathbf{R}^3$. In
particular, $P_\psi (\Delta)=0$ is possible. The reason is
that, in the absence of polarization vectors in
(\ref{p11n}), the Fourier transform  of $\langle \bm{q}
\lambda|\Psi \rangle$ can be an entire function of $\bm{p}$.
Then, according to the Plancherel-Polya theorem~\cite{plancherel}, 
$P_\psi (\Delta)$ may vanish in Borel sets like
$\Delta \in\mathbf{R}^3$. Physically this is necessary 
in order to cope with the absence of photons in $\Delta$.
It is the product of $\tilde{\Psi}^i (\bm{p})$ and
${\epsilon^i (\bm{p},\lambda)}/{\sqrt{2 |\bm{p}}|}$ in (\ref{p11m}) 
that has to be entire. This can be so even if the presence of
$\sqrt|\bm{p}|$ explicitly and in the $\epsilon$'s implies 
that the individual $\tilde{\Psi}^i (\bm{p})$ can not be 
entire functions and, therefore, can not be given 
a probabilistic interpretation. In any case, it is worth 
to recall that the expressions in (\ref{p11m}) are 
for fixed helicity. Thence photon states with fixed helicity
could be localizable in spite of~\cite{galindo} and of theorems 1 and 2 in~\cite{amrein}. This is a first consequence of the until now only hypothesized existence of $\bm{q}$.

The expansion of arbitrary states in $\cal{H}$ in terms of
momentum and helicity eigenstates can be given inverting
(\ref{p11m}) (recall that $\cal{H}$ is the space of {\sl
transverse} states)
\begin{equation}
\tilde{\Psi}^i (\bm{p})=\sqrt{2\, |\bm{p}|} \sum_\lambda
\epsilon^i (\bm{p},\lambda) \langle \bm{p} \lambda|\Psi
\rangle \label{p11u}
\end{equation}
a relation that is customarily used in conjunction with Eq.
({\ref{p1}) to write
\begin{equation}
\tilde{\Psi}^i (\bm{x})=(2\pi)^{-3/2} \sum_{\lambda} \int
\,\,d^3p\,\, e^{i \bm{p}\bm{x}} \, \epsilon^i (\bm{p},
\lambda)\,\, \langle \bm{p} \lambda|\Psi\rangle \label{p11f}
\end{equation}
This is the standard coordinate representation used to
describe the one photon states and --  promoting $ \langle
\bm{p} \lambda|\Psi\rangle $ to the ranks of annihilation
operators -- the photon field. The comparison of
(\ref{p11n}) and (\ref{p11f}) indicates clearly that it is
the presence of the $p$-dependent polarization vector
 in the integrands of (\ref{p11}) and  (\ref{p11f}) that
 forestalls the interpretation of
 $\bm{x}$ as a true position. The decompositions Eqs.
(\ref{p11u}) and (\ref{p11f}) clearly show that the
{$\tilde{\Psi}(\bm{p})$} are transverse (i.e. $\sum_i
p_i\tilde{\Psi}_i(\bm{p})=0$) and the $\tilde{\Psi}
(\bm{x})$ solenoidal (i.e. $\partial
\tilde{\Psi}_i(\bm{x})/\partial x_i)=0$. The interested
reader is referred to ~\cite{iwob} for the use of
(\ref{p11f}) as a photon wave function and to Landau and
Peierls~\cite{landaub} to explore the meaning of the non
local relations between $\tilde{\Psi}(\bm{x})$ and $\langle
\bm{q} \lambda|\Psi\rangle$. It is straightforward to show
that they are
\begin{equation}
\tilde{\psi}^i(\bm{x})=\sum_\lambda \int d^3 q \,
\tilde{V}^i_{\bm{q} \lambda} (\bm{x}) \langle \bm{q} \lambda
|\psi \rangle,\;\; \mbox{and}\; \; \langle \bm{q} \lambda
|\psi \rangle= \int d^3 x \, \tilde{V}^{i*}_{\bm{q} \lambda}
(\bm{x}) \tilde{\psi}^i(\bm{x}) \label{p11g}
 \end{equation}

\section*{Position operator}
In the previous section we have seen that there is room in
the Hilbert space of one photon states for an ordinary
position operator. In fact
\begin{equation}
 \langle \bm{q}' \lambda|\hat{q_i}|\psi  \rangle= q'_i \langle \bm{q}'\lambda|\psi  \rangle
,\;\; \mbox{and}\; \;
 \langle \bm{p} \lambda |\hat{q_i}|\psi  \rangle=i \hbar(\frac{\partial}{\partial p_i})
 \langle \bm{p} \lambda|\psi  \rangle \label{p11h}
 \end{equation}
Here we will analyze the features of this operator as seen
in the space of transverse states. Following
Hawton~\cite{hawtona}, we first identify its structure by
letting it to operate on the eigenfunctions $V_{\bm{q},\lambda}$.
Then we shall show that it qualifies as an appropriate
position operator. Concisely, our first task is to find the
operator $\hat{q}^a,\, a=1,2,3$ whose eigenfunctions are the
$\bm{V}$'s given in Eq. (\ref{p11})
\begin{equation}
 \left\{\hat{q}^a \, \tilde{\bm{V}}_{\bm{q},\lambda}\right\}_i(\bm{p})= q^a \,
 \left\{\tilde{\bm{V}}_{\bm{q},\lambda}\right\}_i (\bm{p}) \label{p12}
 \end{equation}
By using  above the explicit form (\ref{p11}) of
$\tilde{\bm{V}}_{\bm{q},\lambda}$, Hawton got after some
algebra~\cite{hawtona}
 \begin{eqnarray}
\left\{\hat{q}^a \, \tilde{\bm{V}}_{\bm{q},\lambda}\right\}_i (\bm{p})= ( i
\delta_{ij} \nabla^a + (Q^a)_{ij} ) \,
 \left\{\tilde{\bm{V}}_{\bm{q},\lambda}\right\}_j (\bm{p})&\Rightarrow &
 (\hat{q}^a)_{ij}=i \delta_{ij} \nabla^a + (Q^a)_{ij}\label{p13}
\end{eqnarray}
where
\begin{equation}
\nabla^a =\sqrt{2 |\bm{p}|} \frac{\partial}{\partial
p^a}\frac{1}{\sqrt{2 |\bm{p}|}},\;\; \mbox{and}\;\; (Q^a)_{ij} =i
\sum_{\sigma=\theta,\varphi,k} e_i (\bm{p}, \sigma)
\left\{\nabla^a e_j (\bm{p}, \sigma) \right\} \label{p14}
\end{equation}
As is well known, the operator ordering implied by
$\nabla^a$ is necessary to make it self-adjoint with the
measure $d\sigma(\bm{p})$. Notice that the attained operator 
(\ref{p13}) is
independent of the helicity quantum number, being a matrix
in the coordinate indices. By computing the derivatives of
the basis vectors that appear in the definition of $Q^a$,
one obtains the explicit expression of the operator
\begin{equation}
(\hat{q}^a)_{ij}=i \delta_{ij} \nabla^a +\frac{1}{|\bm{p}|}
\left[\bm{e}(\bm{p},k)\wedge \bm{S}_{ij}\right]^a -
\frac{\cot \theta}{|\bm{p}|}
e^a(\bm{p},\varphi) W_{ij}\label{p15a}
\end{equation}
due to Hawton~\cite{hawtona,hawtonb,hawtonc}, who correctly
identified the first two terms as the Pryce position
operator~\cite{pryce}, and the last one as a compensating
term~\cite{iwoc} for the topological photon
phase~\cite{chiao,tomita}. She also realized that, due to the 
last term on the {\sl rhs} of (\ref{p15a}), $\hat{\bm{q}}$
appeared as a set of three components in involution, 
something that previous position operators did not meet.

  A compact expression that summarizes
much of the above, shows explicitly the relation between
this position operator and the spinless one $i \nabla^a$,
and gives a rationale for it, is
\begin{equation}
(\hat{q}^a)_{ij} =i \sum_{\sigma=\theta,\varphi,k} e_i (\bm{p},
\sigma) \nabla^a e_j (\bm{p}, \sigma) \label{p15}
\end{equation}
The sum in (\ref{p15}) includes the longitudinal
polarization $\bm{e}(\bm{p}, k)$ but, in spite of it, $\hat{\bm{q}}$
operates within the transverse subspace $\cal{H}$ overcoming
some queries put forward by Wightman~\cite{wightman}. Using
(\ref{p15}) on an arbitrary function $\bm{\Psi} \in \cal{H}$,
expanded according to (\ref{p11u}), we obtain
\begin{equation}
\left\{ \hat{q}^a \tilde{\bm{\Psi}} \right\}_i (\bm{p}) =\sum_\lambda
\int d^3 p' \tilde{V}^i_{\bm{p}',\lambda}(\bm{p}) i
\frac{\partial}{\partial p'^a} \langle
\bm{p}',\lambda|\bm{\Psi}\rangle= \sqrt{2 \,|\bm{p}|} \sum_\lambda \epsilon_i
(\bm{p},\lambda) i \frac{\partial}{\partial p^a} \langle
\bm{p},\lambda|\bm{\Psi}\rangle \label{p16}
\end{equation}
which shows explicitly the transversality of $\hat{\bm{q}}\cdot
\tilde{\bm{\Psi}}$. We also recall here that
$[\hat{q}^a,\hat{W}]=0$ as can be seen by explicit computation using (\ref{p13}) and the definition of the helicity. Putting all together, it is
possible to use the familiar notations:
\begin{equation}
\hat{q}^a =\sum_\lambda \int d^3 p\,\, |\bm{p} \lambda \rangle i
\frac{\partial}{\partial p^a} \langle \bm{p}, \lambda|, \;\;\;
\mbox{and}\;\;\; \hat{q}^a =\sum_\lambda \int d^3 q\,\, |\bm{q}
\lambda\rangle q^a \langle \bm{q}, \lambda| \label{p17}
\end{equation}
for the position operator in the helicity representation
$\hat{q}^a$. Finally, the wave functions can be interpreted
as in the non relativistic case: given a photon in the state
$\Psi$, $P_\Psi (\bm{p},\lambda)=|\langle \bm{p} \lambda |\Psi
\rangle|^2$ gives the probability of finding it with
helicity $\lambda$ and momentum $\bm{p}$, and $P_\Psi
(\bm{q},\lambda)=|\langle \bm{q} \lambda |\Psi \rangle|^2$  gives the
probability of finding it with helicity $\lambda$ in the
position $\bm{q}$.

 Any arbitrary operator $\hat{\cal{O}}$ defined in the Hilbert
 space $\cal{H}$ can be represented in the two different bases
 introduced above. Using the scalar product (\ref{p2}) and the
 relation (\ref{p11u}) we get:
 \begin{eqnarray}
 \langle \bm{\Phi}, \hat{\cal{O}}\; \bm{\Psi} \rangle &=& (2 \pi)^{-3/2} \int
 d\sigma(\bm{p})\, \sum_{ij}\, \tilde{\Phi}^*_i (\bm{p}) \hat{\cal{O}}_{ij}
  \tilde{\Psi}_j (\bm{p})  \\
  &=& (2 \pi)^{-3/2} \int d^3 p \sum_{\lambda\,\lambda'} \langle
  \bm{\Phi} |\bm{p}\, \lambda \rangle\, \hat{\cal{O}}_{\lambda\,\lambda'}
  \langle \bm{p}\, \lambda  | \bm{\Psi} \rangle \label{p18}
  \end{eqnarray}
so that there is a well defined relation between the
operator's expressions in both bases:
\begin{equation}
\hat{\cal{O}}_{\lambda\,\lambda'}(\bm{p}, \partial/\partial \bm{p}) =
\frac{1}{\sqrt{2 |\bm{p}|}}\sum_{ij} \epsilon^*_i (\bm{p},\lambda)
\hat{\cal{O}}_{ij}(\bm{p}, \partial/\partial \bm{p}) \epsilon_j
(\bm{p},\lambda') \sqrt{2 |\bm{p}|} \label{p19}
\end{equation}
Applying this to the canonically conjugate momenta
$\tilde{p}^a_{ij}=\delta_{ij} p^a$ and position (\ref{p15})
operators, we get them in the helicity basis as
\begin{equation}
\hat{p}^a_{\lambda \lambda'}=\delta_{\lambda \lambda'}
p^a;\,\,\, \mbox{and}\,\,\, \hat{q}^a_{\lambda
\lambda'}=\delta_{\lambda \lambda'} i
\frac{\partial}{\partial p^a} \label{p20}
\end{equation}
We have thus recovered the old Heisenberg-Weyl quantization
rules as anticipated in (\ref{p11h}). They are valid in the
helicity basis and only in it because, due to helicity
conservation, additional terms appear in other frames to
compensate for the effects of $\partial/\partial \bm{p}$ on the
momentum dependent polarizations. As said above,  the
meaning of the derivative in the helicity basis is that of a
covariant derivative. In fact, it is related to the standard
covariant derivative introduced in ref~\cite{iwoc} to
account for the topological phase~\cite{chiao,tomita} of the
photon.
\begin{equation}
\widehat{(i\nabla^a)}_{\lambda \lambda'}= \delta_{\lambda
\lambda'} \left[  i \frac{\partial}{\partial p^a} + \lambda
D^a (\bm{p})\right] \label{p21}
\end{equation}
where $D^a (\bm{p})=\cot \theta e^a(\bm{p},\varphi)/|\bm{p}|$. Notice, by the
way, that the helicity operator $\hat{W}_{ij}$ transforms to
the helicity quantum number $\lambda$ in the transformation
from the spin basis onto the helicity basis. The spin matrix
also undergoes a similar transformation
\begin{equation}
 \hat{S^a}_{\lambda \lambda'} =\delta_{\lambda
\lambda'} \lambda e^a(\bm{p},k) \label{p22}
\end{equation}
The transformation to the helicity basis analyzed above
$\hat{{\cal O}}_{ij} \rightarrow \hat{{\cal O}}_{\lambda
\lambda'}$ should not be mistaken with the similarity
transformation to the photon frame, namely:
\begin{equation}
\hat{{\cal O}}^a_{\sigma \sigma'}=\sum_{ij} e_i(\bm{p},\sigma)
\hat{{\cal O}}^a_{ij} e_j(\bm{p},\sigma'), \,\,\,
\mbox{with}\,\,\, \{
\sigma,\sigma'\}=\{k,\theta,\varphi\}\label{p24}
\end{equation}
This is simply the rotation from the fixed cartesian axes to
the axes lying along the unit vectors $\bm{e}(\bm{p},\sigma)$. A most
striking case occurs for the spin matrix that, by (\ref{p24}),
becomes $\hat{S}^a_{\sigma \sigma'}=i \sum_{\sigma''}
\epsilon_{\sigma \sigma' \sigma''} e^a(\bm{p},\sigma'')$. Notice
the difference with (\ref{p22}):
\begin{equation}
\left[\hat{S}^a,\hat{S}^b\right]_{\sigma \sigma'}= i \sum_c
\epsilon_{a b c} \hat{S}^c_{\sigma \sigma'}\,\,\,
\mbox{while} \,\,\,
\left[\hat{S}^a,\hat{S}^b\right]_{\lambda \lambda'}=0
\label{p25}
\end{equation}
The vanishing of the second commutator is of no surprise as
there are no remains of the spin matrices in the helicity
representation. The dimension of the spin space is 3, while
the helicity basis is a sum of two one-dimensional
representations. Even after putting together the two parity
related representations (1,0) and (0,1) to have both
helicities, the would-be  helicity 0 eigenstate is outside
the representation space. The relation of these facts with
gauge invariance was discussed~\cite{weinbergb,weinbergc} a long
time ago. From the geometric point of view, the
transformation $\hat{{\cal O}}_{ij} \rightarrow \hat{{\cal
O}}_{\lambda \lambda'}$ is a projection from $\mathbf{R}^3$
onto the transverse subspace spanned by $\bm{e}(\bm{p},\theta)$ and
$\bm{e}(\bm{p},\varphi)$, along with the appropriate label
rearrangements. As a result, the  operator products must be
handled with caution: in general their transform shall not
coincide with the product of the operators' transforms. We
saw this in Eq. (\ref{p25}) for the spin matrices. The same
occurs for the angular momentum and the boost operators and,
in general, for all the operators that pull the states out
of the transverse space. This sometimes led to define the
helicity representation through a non singular
transformation applied to the spin representation:
\begin{equation}
\hat{\cal{O}}'_{ij}(\bm{p}, \partial/\partial \bm{p}) =
\frac{1}{\sqrt{2 |\bm{p}|}}\sum_{rs} R^{-1}_{ir}(\bm{p})
\hat{\cal{O}}_{rs}(\bm{p},
\partial/\partial \bm{p}) R_{sj}(\bm{p}) \sqrt{2 |\bm{p}|}
\label{p26}
\end{equation}
where $R(\bm{p})$ can be any of the rotations from the standard
momentum $(0,0,|\bm{p}|)$ to $\bm{p}$. Denoting by $\bm{e}(\sigma),
\sigma=1,2,3$ three unitary vectors along the fixed axis
\begin{equation}
e_i(\bm{p},\sigma)=R_{ij}(\bm{p}) e_j(\sigma) \label{p27}
\end{equation}
Notice that the intrinsic arbitrariness in $R$, due to the
invariance of the standard momentum under rotations around
it, is removed by the choice of fixed vectors $e_i(\sigma) $
in (\ref{p27}). Note also that the helicity representation
operators are obtained by projecting (\ref{p26}) on the
fixed helicity basis:
\begin{equation}
\hat{\cal{O}}_{\lambda \lambda'} = \,\,\sum_{i j}\,\,
\epsilon^*_i(\lambda) \,\,\hat{\cal{O}}'_{ij} \,\,
\epsilon_j(\lambda')  \label{p27a}
\end{equation}
where, according to (\ref{p4}) and (\ref{p27}),
$\epsilon_i(\lambda)=\frac{\lambda}{\sqrt{2}}
\{\epsilon_i(1)+i\lambda \epsilon_i(2)\}$. Needless to say
that (\ref{p27a}) is invertible only within the subspace
orthogonal to the standard momentum.

The application of the
above results to the specific case of the electromagnetic
field completes the standard picture of the one photon state
with additional, position dependent, information. A photon
in a state $\bm{A}$ can be given as
\begin{equation}
A^i (\bm{x})=(2\pi)^{-3/2} \sum_{\lambda} \int  \,\,\frac{d^3
p}{\sqrt{2\, |\bm{p}|}}\,\, e^{i \bm{p}\bm{x}} \, \epsilon^i (\bm{p}, \lambda)\,\,
\langle \bm{p} \lambda|\,\,\bm{A}\,\,\rangle \label{p27aa}
\end{equation}
or in the alternative form
\begin{equation}
A^i(\bm{x}) =\sum_\lambda \int dq V^i_{\bm{q},\lambda}(\bm{x}) \,\, \langle
\bm{q}\,\, \lambda|\,\,\bm{A}\, \,\rangle \label{p27e}
\end{equation}
this reinforces the interpretation of $\hat{\bm{q}}$ as a
position operator and of $\bm{V}_{\bm{q} \lambda}(x)$ (given explicitly in (\ref{p11}) as the
configuration space amplitude of the photon state localized
at $\bm{q}$ with helicity $\lambda$.

Note that we could add  zero components to the polarization
vectors to form a fourvector-like object $A_\mu (\bm{x})$.
However, as  shown in (\ref{p4a}) this object does not
transform as a fourvector, but inhomogeneously as a
connection:
\begin{equation}
{\cal U}[\Lambda] A_\mu (x,\lambda) {\cal U}[\Lambda^{-1}]=
\Lambda_\nu^\mu\,\, (2 \pi)^{-3/2} \int \frac{d^3 p}{\sqrt{2
p}} \,\,e^{-i p \Lambda x} \,\,\{\epsilon_\nu (p,\lambda)-
p_\nu\,\, \Omega(p,\lambda;\Lambda)\}\,\,\, \langle p
\,\lambda |\,\,A \rangle \label{p27b}
\end{equation}
A tensor-like covariant object for the photon can be defined
as
\begin{equation}
F_{\mu \nu}(\bm{x})=(2\pi)^{-3/2} \sum_{\lambda} \int
\,\,\frac{d^3 p}{\sqrt{2\,|\bm{p}|}}\,\, e^{i \bm{p}\bm{x}} \,
i\,\,\{p_\mu\, \epsilon_\nu (\bm{p}, \lambda)- p_\nu\,
\epsilon_\mu (\bm{p}, \lambda)\}\,\, \langle \bm{p}
\lambda|\,\,A\rangle \label{p27c}
\end{equation}
This contains the same information that $A_\mu$, but
transforms as an antisymmetric tensor; its components 
constitute the electric and magnetic fields. In the next 
section we shall use (\ref{p27c}) to construct and solve the
photon Maxwell equations in vacuum.

\section*{Time operator}
The  Galileo boost operator  in the NR representations for
elementary systems has only three components $\hat{G}_i,
i=1,2,3$. Each of them is associated to a component of the
position operator. There is no room in the representation
space for an additional boost associated to the time. In
fact, time is invariant under Galileo boosts. Most likely,
this is hidden among the reasons~\cite{pauli,mandelstamm,fock,paul,aharonov,kijowski,
buscha,buschb,hilgevoorda,hilgevoordb} for the
difficulty of finding a time operator that
behave just like the other position operators, something
tempting in the study of individual particles. As explained in~\cite{hilgevoordb}, this is a
misleading approach to the role of time in
quantum mechanics and its use led to several dead ends in
the past. A question to be taken into account is that time
translations are given in terms the Hamiltonian $H$ which is
not independent, but is fixed as a function of the other
operators in the representation by the mass shell condition.
The same is to be expected for a time operator, instead of
the addition of a new independent component to the position,
a given function of the basic operators, whose explicit form
depends on the very properties of the representation.

Leaving aside the far-reaching but still out of reach question 
of the status of
Newtonian time in quantum mechanics, and the search of its corresponding
operator  -- if any -- it is possible to identify~\cite{grot,reisenberger}
 time-like properties of quantum systems. The
simplest of these appears for free elementary particles in
the form of the time of arrival at a fixed position
$\bm{X}$, a deceptively simple property whose analysis in
quantum mechanics is very delicate and full of traps.
Classically it is a derived quantity, a function of the
initial position and momentum whose values are either
``never" (when $\bm{X}$ is out of the particle trajectory),
or a real number that solves the equations of motion for $t$
as a function of $\bm{X}$. Notice that some kind of
integrability is implicit for this classical notion to work
properly~\cite{leonc}. Quantization requires the splitting of
the Hilbert space into two orthogonal subspaces, one that
contains the never detected states, and other that contains
the eventually detected ones. The time of arrival acts
within this last subspace~\cite{grot,leona}, where it is
represented by an operator which is not self-adjoint but
only maximally symmetric. Its eigenstates are not
orthogonal, so that instead a projector valued measure, the
statistics of the times of arrival at a detector can only be
analyzed in terms of positive operators~\cite{giannitrapani}.
Below, I shall show how this works for the photon.

First, we shall study the time dependence hidden 
in the bracket $\langle \bm{p}
\lambda|\,\,A(t)\rangle$.  Taking into account that
$p^2=0$, and that $ p\, \epsilon(\bm{p},\lambda)=0$, we get from
(\ref{p27c}) the  equations $\partial^\mu F_{\mu \nu}=0$.
These cover the full set of Maxwell equations due to the
duality relation among null fields in vacuum~\cite{weinbergc}:
\begin{equation}
F^{\mu \nu}(x,\lambda)=i\,\,\frac{\lambda}{2}\,\,
\epsilon^{\mu \nu \rho \sigma} \,\, F_{\rho \sigma}
(x,\lambda)\Rightarrow \left\{ \begin{array}{rcl} F_{ij} & =
& \epsilon_{ijk}
(E_k+i\lambda\,B_k)\\&&\\
 F_{0k}&=& -i \lambda (E_k+i\lambda\,B_k)
 \end{array} \right.
 \label{p27d}
\end{equation}
We now have two equations $\partial^i F_{i0}=0$ and
$\partial^0 F_{0j}+\partial^i F_{ij}=0$. Calling
$\bm{F}(\bm{x}, \lambda)=\bm{E}(\bm{x}, \lambda)+i \lambda
\bm{B}(\bm{x}, \lambda)$, the first equation reduces to the
divergenceless condition $\bm{\nabla}\bm{F}(\bm{x},
\lambda)=0$ and the second to the Schr\"odinger equation~\cite{pryce,cooka,inagaki,iwob,sipe}:
\begin{equation}
i \frac{\partial \bm{F}(\bm{x}, \lambda)}{\partial t}=
\lambda c \bm{\nabla}\wedge \bm{F}(\bm{x},
\lambda)\label{q10}
\end{equation}
Recalling the expression for the spin matrix $(S^j)_{ik}=i
\epsilon_{ijk}$, this equation can be given a more
transparent form:
\begin{equation}
i\frac{\partial F_i(\bm{x}, \lambda)}{\partial t} = \lambda
(\bm{S} \bm{p})_{ij}F_j(\bm{x}, \lambda)\label{q11}
\end{equation}
where $\bm{p}=-i \bm{\nabla}_{\bm{x}}$. Using now
(\ref{p27c}) in the definition of $\bm{F}$ we obtain
\begin{equation}
\bm{F}(\bm{x}, \lambda)=(2\pi)^{-3/2} \int d^3 p \sqrt{2
|\bm{p}|}\, i\,e^{i \bm{p}\bm{x}}
\bm{\epsilon}(\bm{p},\lambda)\,\, \langle \bm{p}
\lambda|A(t)\rangle \label{q12}
\end{equation}
therefore, $(\bm{S}\bm{p})_{ij}F_j=\lambda |\bm{p}| F_j$ and
hence,
\begin{equation}
i\frac{\partial \bm{F}(\bm{x}, \lambda)}{\partial t}=c\,\,
|\bm{p}|\,\,\bm{F}(\bm{x}, \lambda) \label{q13}
\end{equation}
Therefore, the photon Hamiltonian is $H= c|\bm{p}|
\delta_{\lambda \lambda'}$ in the helicity basis.

Due to the simple expression  of the photon Hamiltonian, the
position operator in the in the helicity basis evolves quite
simply:
\begin{equation}
\frac{d\hat{q}^a}{d t}=i [\hat{q}^a,c
\sqrt{\hat{\bm{p}}^2}]=c\,\,
\frac{\hat{p}^a}{|\hat{\bm{p}}|},\,\,\,
\,\,\,\frac{d\hat{p}^a}{d t}=0\,\,\,\,\, \Rightarrow \,\,\,
\hat{q}^a (t)=\hat{q}^a + c\,\,
\frac{\hat{p}^a}{|\hat{\bm{p}}|}\, t,\,\,\,
\hat{p}^a(t)=\hat{p}^a \label{p28}
\end{equation}
In the first of these Eqs.  we displayed explicitly the
hamiltonian obtained from the mass shell condition
$H(p)=c\sqrt{\bm{p}^2}$, but then we continue to denote it
by $|\bm{p}|$ as before. $\hat{q}^a$ and $\hat{p}^a$ are the
position and momentum operators of the photon at $t=0$,
while $\hat{q}^a(t)$ is the position operator at a time $t$
(we are working in the Heisenberg picture). All this is
trivial and prompts to the seemingly innocent question: When
does the photon arrives at a position $\bm{x}$? This is the
alternative to the standard quantum mechanical question:
Where is the photon at time $t$? Much in the same way that
the question ``where?" brings a position operator whose
probability distribution is used to ascertain {\sl at what
place}?, the question ``when?" demands the introduction of a
time operator, with the meaning of time of arrival (ie. {\sl
at what instant}?), to close the logical loop. Several
obstructions prevent the existence of such an operator in
the standard formulation of quantum mechanics. The earliest
one, which turned out to be the strongest, was formulated in
the early days of quantum theory by Pauli~\cite{pauli}: The
simultaneous existence of unitary representations for both,
energy and time conjugate operators, is incompatible with a
bounded hamiltonian, opening up at least a sort of infrared
catastrophe (see however~\cite{hilgevoordb}). This led to renounce to the self-adjointness of
the time operator, keeping its maximal symmetry only (i.e.
${T^*}^\top =T$). In these conditions, time eigenstates are
no longer orthogonal (unless some steering regularization is
used~\cite{grot}) and, instead of the traditional projector
valued measure, it is necessary to turn to a positive
operator valued measure~\cite{giannitrapani} to interpret the
formalism. The interested reader is referred to the
review~\cite{muga} to get an account of these and related
issues, and references to the original literature.

Generally, the search of a time of arrival operator has been
undertaken in the case of one space dimension. The present
author studied the case of the free particle in three space dimensions~\cite{leona} concluding that the
detected states are confined within a subspace of the whole Hilbert space. Outside it is the realm of no
detection, that is, of those states for which the time of
arrival is ``never." In classical terms, they miss the
detector, whose efficiency -- a different question -- is 
assumed to be
$1$.  On physical grounds, detection requires a constraint:
that the particle momentum is parallel to the line joining
${\bm{q}}$ with the arrival (detector) position $\bm{z}$.
This vector constraint and the free particle Hamiltonian
form a first class system. The use of this formulation made
possible the obtention in~\cite{leona} of a very simple
solution for the time of arrival in three dimensional space,
basically an extension of the 1-D results.

A time operator for photons requires of a 3-D setting. Not
only because the transverse-vector character of the
electromagnetic field  stripes the 1-D approach of
credibility. Also from the practical side: we shall analyze
later on the arrival of photons through inhomogeneous media,
where direction changes will in general take place, whose
description needs of more than the mere distance covered.
Therefore, we shall examine the first class constrained
system that evolves in the detected subspace  as the first
step towards the time of arrival of photons.

The task can be formulated in very simple terms: If $\bm{z}$ is
the detector  position, at what time $t$ is $\hat{q}^a
(t)=z^a$? In other words: How to invert the equation
(\ref{p28}), namely $z^a=\hat{q}^a + c\,\,
({\hat{p}^a}/{\hat{\bm{p}}})\,\, t$, to get $t$? Several comments
are in order here. First, we are promoting $t$ to the
category of a q-number, while demoting $\hat{\bm{q}}(t)$ to a
given, external, parameter. This is the very task to
accomplish to define a time operator. Second, the evolution
equations that we have to invert to obtain the time of
arrival is the set (\ref{p28}) of three equations, one per
component, depending on a unique parameter $t$. To be
compatible, they have to satisfy the constraint:
\begin{equation}
{\mathbf L}_a (\bm{z}) = \epsilon_{a b
c}(\hat{q}_b-z_b)\,\,\hat{p}_c = 0\label{p29}
\end{equation}
To quantize this constrained system we borrow from the method
of Dirac~\cite{dirac}. Classically, the constraint
guarantees that the orbital angular momentum of the particle
is $\bm{z}\wedge\bm{p}$, so that $\bm{z}$ is a point  of its
trajectory. The total Hamiltonian formed by adding the
constraints to the original one is:
\begin{equation}
H_\top(\bm{z})=c \sqrt{\bm{p}^2}\,\,+\,\,\mu_a 
{\mathbf L}_a (\bm{z})
\label{p30}
\end{equation}
where $q^a$ and $p^a$ are the dynamical variables to become
operators after quantization, the $\mu$'s are Lagrange
multipliers, and $\bm{z}$ is an external vector parameter
corresponding to the detection position. The system is first
class:
\begin{equation}
\{{\mathbf L}_a (\bm{z}),\,\,{\mathbf L}_b (\bm{z})\}=\epsilon_{a b
c}\,\, {\mathbf L}_c (\bm{z}),\;\;\; \{{\mathbf L}_a
(\bm{z}),\,\,H_\top(\bm{z})\}=\epsilon_{a b c}\,\, \mu_b\,\,\,
{\mathbf L}_c (\bm{z}) \label{p31}
\end{equation}
where $\{,\}$ indicate Poisson brackets as this is still
classical dynamics. The evolution of the constraints does
not produce additional (secondary) constraints. Hence, the
Hamiltonian (\ref{p30}) is enough to account for the
evolution of the constrained system.

When quantizing the system, the vanishing of the constraint
translates into a kind of subsidiary condition:
\begin{equation}
\left(\hat{{\mathbf L}}_a (\bm{z}) \tilde{\Phi}\right)_i 
(\bm{p})=0
\label{p32}
\end{equation}
The set of vectors of the Hilbert space $\tilde{\Phi}_i (\bm{p})$
that satisfy the above equation form the subspace ${\cal
H}_{\bm{z}}$ of the states that could, eventually, be detected at
$\bm{z}$. On the other hand, using (\ref{p16}), Eq. (\ref{p32})
can be written as:
\begin{equation}
\epsilon_{a b c} \sqrt{2 |\bm{p}|} \sum_\lambda \epsilon_i
(\bm{p},\lambda)\,\, \{i \frac{\partial}{\partial p_b} - z_b
\}\,\, p_c \,\,\langle \bm{p} \lambda |\Phi \rangle=0
\label{p33}
\end{equation}
whose solution is
\begin{equation}
\langle \bm{p} \lambda |\Phi_{\bm{z}} \rangle = e^{-i \bm{p} \bm{z}}\,\,\,
\Phi(|\bm{p}|,\lambda,\bm{z}) \label{p34}
\end{equation}
Note that the functions $\Phi$ may depend on the modulus of
the momentum (but not on its direction), on the helicity
and, possibly, on the detection point $\bm{z}$. However, this
last dependence has to be switched off to maintain
translational invariance.

We now proceed to invert Eq. (\ref{p28}). First of all we
write down the action of the position operator on the
detected subspace:
\begin{equation}
\hat{q}^a \,\,\,\langle \bm{p} \lambda |\Phi_{\bm{z}} \rangle = \,\,
e^{-i \bm{p} \bm{z}}\,\,\,\{z^a+i\frac{p^a}{|\bm{p}|}
\,\frac{\partial}{\partial |\bm{p}|}\}\,\,\, \Phi(|\bm{p}|,\lambda)
\label{p35}
\end{equation}
Hence, with $z^a$ a parameter and $\hat{t}(\bm{z})$ an
operator, Eq. (\ref{p28}) reads:
\begin{equation}
z^a=\hat{q}^a + c\,\, \frac{p^a}{|\bm{p}|}\,\,\, \hat{t}(\bm{z}),\,\,\,
\Rightarrow\,\,\, \,\,e^{-i \bm{p} \bm{z}}\,\, i\,\,\frac{p^a}{|\bm{p}|}
\,\frac{\partial}{\partial |\bm{p}|}\,\,
\Phi(|\bm{p}| ,\lambda)+c\,\,\frac{p^a}{|\bm{p}|}\,\,\,
\hat{t}(\bm{z})\,\,\,\, e^{-i \bm{p} \bm{z}}\,\,\Phi(|\bm{p}|,\lambda)=0
\label{p36}
\end{equation}
This equation has to be valid whatever the function $\Phi$
chosen. This serves to define $\hat{t}(\bm{})$: It is precisely
the operator that transforms (\ref{p36}) into an identity in
${\cal H}_{\bm{z}}$, that is:
\begin{equation}
\hat{t}(\bm{z}) \approx -i\,\, e^{-i \bm{p}
\bm{z}}\,\,\,\frac{\partial}{\partial |\bm{p}|}\,\,\,\,e^{i \bm{p}
\bm{z}},\,\,\,\,\hat{t}(\bm{z}) = -i\,\, e^{-i \bm{p}
\bm{z}}\,\frac{1}{|\bm{p}|}\,\,\frac{\partial}{\partial |\bm{p}|}\,\,|\bm{p}|
\,\,\,e^{i \bm{p} \bm{z}} \label{p37}
\end{equation}
The symbol $\approx$  at the left indicates equal up to
operator ordering, something that we fix  by the condition
that $\hat{t}$ be maximally symmetric in the integration by
parts with the measure $d^3p$ of the $|\bm{p} \lambda \rangle$
basis. This produces the operator defined at the right hand
side, that we shall use as the time of arrival operator in
what follows. It depends parametrically on $\bm{z}$ in as much
the same way as the operators depend parametrically on $t$.
Incidentally, this makes us recall that we are working in
the Heisenberg picture at $t=0$. It is straightforward to
show that, when some time  $t_0$  has elapsed, the time
operator shifts to $\hat{t}(\bm{z},t_0)=\hat{t}(\bm{z})-t_0$, and that
the arrival occurs at a time $t_{\bm{z}}$ such that
$\hat{t}(\bm{z},t_{\bm{z}})=0$.

The eigenfunctions $\langle \bm{p} \lambda| t \bm{z}\rangle$ of
$\hat{t}(\bm{z})$ obtained by solving the eigenvalue equation
$\hat{t}(\bm{z})\,\,\langle \bm{p} \lambda| t \bm{z}\rangle= t
 \langle \bm{p}
\lambda| t \bm{z}\rangle$ are proportional to $\exp{i(H t-
\bm{p}\bm{z})}/|\bm{p}|$.
Due to the fact that
$\lim_{|\bm{p}|\rightarrow 0} (|\bm{p}| \,\,\langle\bm{p} \lambda | t \lambda
\rangle) \neq 0$, being $|\bm{p}|=0$
 the lower bound of the Hamiltonian, the operator (\ref{p37})
  can not be self-adjoint.
This is similar to the case of the radial momentum $p_r=i
r^{-1} (\partial/\partial r) r$ in three space dimensions
(but notice that this does not prevent us from using
$H=p_r^2/2m \,\,+\,\, l(l+1)/2m r^2$ as Hamiltonian; what is
done in this case is to restrict the behavior of the wave
function at the boundary $r=0$). The time operator commutes
with the helicity so, taking into account that $\langle \bm{p}
\lambda | \bm{q} \lambda' \rangle= \delta_{\lambda \lambda'} (2
\pi)^{-3/2} \exp{(-i \bm{p}\bm{q})}$, we could write
\begin{equation}
\langle\bm{p}\,\,\lambda'|t\,\,\lambda; \bm{z} \rangle =\langle
\bm{p}\,\,\lambda'|\,\,\,\frac{1}{H} e^{i H t}\,\,\,|\bm{z}\,\,\,
\lambda \rangle\,\,\, \Rightarrow \,\,\, |t \,\,\lambda;
\,\,\bm{z}\rangle = \frac{1}{H} e^{i H t}\,\,\,|\bm{z} \,\,\lambda
\rangle \label{p38}
\end{equation}
This notation  is highly symbolic mainly due to the fact
that $\bm{z}$ is just an external parameter belonging to the
experimental set-up, the observer's will, etc,  so that
there is nothing like $\bm{z}$ among the properties of the
particle. However, the above expression is correct if one
considers that $|\bm{z} \,\,\lambda \rangle$ is the eigenket of
the particle's position operator with eigenvalue $\bm{z}$.
Writing the detected states (\ref{p34}) in the same form may
through some light on the meaning of the notation:
\begin{equation}
\langle \bm{p}\,\,\lambda \,\,|\Phi_{\bm{z}}\rangle =\langle
\bm{p}\,\,\lambda \,\,|\,\Phi(H,\,\,\lambda))|\bm{z} \,\,\,\lambda
\,\,\rangle \label{p39}
\end{equation}
This is the effect of the subsidiary condition (\ref{p32}):
it projects on the detector position $\bm{z}$, keeping only that
part of the state that is in $s$-wave relative to $\bm{z}$ (hence
the $\Phi$ dependence on $|\bm{p}|$ alone). The lack of
completeness this produces on the time operator, and the
associated interpretation, was discussed with some detail in
ref.~\cite{leona} for the relativistic massive spinless particle. I shall repeat here the two main
results tailored to the massless, helicity $\pm 1$, photon case:\\
1. The time eigenstates are not orthogonal:
\begin{eqnarray}
\sum_{\lambda'} \int d^3 p && \langle
t\,\,\lambda;\,\,\bm{z}|\,\,\bm{p}\,\,\lambda'
\rangle
\langle\,\,\bm{p}\,\,\lambda'|\,\,t'\,\,\lambda";\,\,\bm{z}\rangle \nonumber\\
=&& \delta_{\lambda\,\,\lambda"}\,\,\,
\frac{1}{2\,\pi^2}\,\,\, \int_0^\infty d|\bm{p}|
\,\,\,e^{i|\bm{p}|(t'-t)}= \delta_{\lambda\,\,\lambda"}\,\,\,
\frac{1}{2\,\pi^2}\,\,\,\frac{i}{t'-t+i\,\,\epsilon}
\label{p40}
\end{eqnarray}
2. The basis is complete only within ${\cal H}_{\bm{z}}$. In other
words, the projection over states orthogonal to the detector
position is excluded from the decomposition of the identity
in terms of time eigenstates:
\begin{eqnarray}
\sum_{\lambda'} \int_{-\infty}^{+\infty}  dt\,\,\, \langle
\bm{p}\,\,\lambda|\,\,t\,\,\lambda';\,\,\bm{z}
\rangle\,\,\langle\,\,t\,\,\lambda';\,\,\bm{z}|\,\,\bm{p}"\,\,\lambda"\rangle
&=& \delta_{\lambda\,\,\lambda"}\,\, \frac{2
\pi}{|\bm{p}|^2}\,\,\delta(|\bm{p}|-|\bm{p}"|)\,\,\langle
\bm{p}\,\,\lambda|\bm{z}\,\,\lambda\,\,\rangle\,\,
\langle\,\bm{z}\,\,\lambda\,\,|\bm{p}"\,\,\,\lambda"\,
\rangle\,\,\, \nonumber \\
\Rightarrow\,\,\,\sum_{\lambda'} \int_{-\infty}^{+\infty}
dt\,\,\, \langle \bm{p}\,\,\lambda|\,\,t\,\,\lambda';\,\,\bm{z}
\rangle\,\,\langle\,\,t\,\,\lambda';\,\,\bm{z}|\,\,\Phi_{\bm{z}}\rangle
&=& \langle \bm{p}\,\,\lambda|\,\,\Phi_{\bm{z}}\rangle \,\,\,\,\,
\forall \,\,\Phi_{\bm{z}} \,\in\,{\cal H}_{\bm{z}} \label{p41}
\end{eqnarray}

We recall that due to the non-orthogonality of the states,
the spectral decomposition of the operator
\begin{equation}
\hat{t}(z) =\sum_\lambda \,\,\,\int dt\,\,\,t\,\,\,
|\,\,t\,\,\lambda;\,\,\bm{z}
\rangle\,\,\langle\,\,t\,\,\lambda;\,\,\bm{z}| \label{p42}
\end{equation}
defines a positive operator valued measure only (not a
projector valued one) that can be used to give the
probability that the arrival of the state $\psi$ at the
detector  occurs in the time $t$ as $P_\psi(t;\bm{z})=
\sum_\lambda\,\,|\langle\,\,t\,\,\lambda;\,\,\bm{z}|\,\,\psi
\,\,\rangle|^2$.
This really is a probability density, something  not
reflected on the notation for the sake of simplicity. Note
also that, due to the lack of completeness, the probability
of eventually arriving at $\bm{z}$ (in any time) $P_\psi (\bm{z})=
\int dt\,\,P_\psi(t;\bm{z})\leq 1$, and may be zero for states
$\psi \notin {\cal H}_{\bm{z}}$. Finally, the mean value of the
time of arrival at $\bm{z}$ of a particle in the state $\psi$ is
$t_\psi (\bm{z})=\langle\,\, \psi\,\,|\hat {t}(\bm{z})|\,\,\psi
\,\,\rangle/P_\psi (\bm{z})$, this excludes counterfactuals (the
case $P_\psi (\bm{z})=0$) as it should.

\section*{Conclusions}
Photons are eventually detected through their interaction with matter. Gauge invariance singles out the minimal coupling of the potentials to the currents $J_\mu (\bm{x},t)$ of additively conserved quantum numbers (the electric charge in this case). The coupling to Pauli like currents (e. g. $\partial^{\nu}\bar{\psi}\sigma_{\mu\nu} \psi$) would never produce the finite amplitudes for absorbing or emitting soft photons observed experimentally. These facts and their implications have been thoroughly analyzed by S. Weinberg~\cite{weinbergc}and other authors. Of course, by means of a canonical transformation~\cite{power,woolley} the minimal coupling interaction can be cast into a multipolar form. We will take into account this structure of the interactions when discussing the detection of photons. Assume for simplicity the case of broad-band photodetectors for which the counting rates are given by the energy density of the field
$\langle \psi |\hat{\bm{E}}^{\dagger}(\bm{x},t) \cdot \hat{\bm{E}}(\bm{x},t)| \psi \rangle$. Then, the probability that the state $\psi$ be localized (in the plain sense of being detected by a detector) in a neighborhood $\Delta$ around $(\bm{x},t)$ is given in terms of the fraction of the total energy that is within $\Delta$. 

Absolute localization of quantum mechanical systems~\cite{wightman} -that is, the condition that the probability of finding the system out of some finite volume vanish- is such a strong condition that it violates causality~\cite{hegerfeldta,hegerfeldtb}. In other words, any free particle initially confined in a finite volume, continues in it forever, or immediately spreads to infinity. This result applies to free relativistic and non relativistic particles, to complex systems, in the presence of interactions, etc. Surprisingly the only requirement for this is that the system Hamiltonian be bounded from below. This prompts the question of what is the maximal degree of localization to be expected for a particle. An important clue~\cite{hegerfeldtc} is that -- even if the localization outside the finite volume is not absolute, but exponentially bounded tails are permitted -- the probability spreads  out to infinity faster than with any finite propagation speed.

The limits to photon localization have been strengthened recently~\cite{iwoc} based on simple physical requirements to be satisfied by one photon states. Basically, these are: a) That $\langle \bm{p} \lambda | \psi \rangle$ can be given a probabilistic interpretation (something that we discussed below (\ref{p11z})) and b) That the Hamiltonian be bounded from below. Then, the Paley-Wiener Theorem VIII~\cite{paley} says that the fall-off of the photon wavefunction $\langle \bm{q} \lambda | \psi \rangle$ as $|\bm{q}|  \rightarrow \infty$ is slower than $\exp (-a |\bm{q}|^r)$, where $a,r$ are positive constants and $r<1$. As the physical requirements noted above apply to all types of particles, the same occurs to the limit of almost exponential localization: it applies to all kind of particles. This puts photons at the same level than the other particles in what refers to localization.
A recent analysis of spontaneous emission from excited atoms~\cite{chan} has shown the possibility of producing entangled atom-photon states where the photon wave packets have Gaussian tails. This explicit breaking of the barrier of exponential localization is a product of the entangled final state. 

These results shall likely find their application in the field of quantum information, and shall promote new developments in quantum optics. Some necessary tools like  good position~\cite{hawtona} and time of arrival operators and their associated probabilistic interpretations are provided in this paper. We are completing a detailed analysis of the application of these tools to the tunneling through photonic band gaps, to HOM~\cite{hong} interferometry and entanglement~\cite{chan,can}, and to the superluminal propagation detected in several experiments~\cite{berkeley1,vienna,berkeley2}.

\begin{acknowledgments}
I want to thank J. Julve and F.J. de Urr\'{\i}es and L. Lamata for useful discussions and valuable comments on the paper.
This work was partially supported by the Spanish Ministry of Science 
and Technology under the project BMF 2002-00834.
\end{acknowledgments}


\end{document}